\documentclass[showpacs,pra,twocolumn,aps,amsmath,amssymb,unsortedaddress]{revtex4}
\usepackage{graphicx}
\usepackage{dcolumn}
\usepackage{amsmath}
\usepackage{times}

\begin{document}

\draft

\title{Green's operator for Hamiltonians with Coulomb plus polynomial potentials}
\author{E.\ Kelbert, A. Hyder, F.\ Demir, Z.\ T.\ Hlousek and Z.\ Papp}
\affiliation{ Department of Physics and Astronomy, California State
University, Long Beach, California 90840 }

\date{\today}

\begin{abstract}\noindent
The Hamiltonian of a Coulomb plus polynomial potential on the Coulomb-Sturmian basis
has an infinite symmetric band-matrix structure. A band matrix can always be considered as a block-tridiagonal matrix. So, the corresponding Green's operator can be given as a matrix-valued continued fraction. As examples, we calculate the Green's operator for the Coulomb plus linear and quadratic potential problems and determine the energy levels.
\end{abstract}
\pacs{03.65.Nk  }

\maketitle

\section{Introduction}

Coulomb plus polynomial potentials,
\begin{equation}\label{v1}
v(r)=\sum_{i=-1}^{k} a_{i}r^{i}=a_{-1}/r+a_{0}+a_{1}r + a_{2}r^{2}+\cdots~,
\end{equation}
are often used to model various physical phenomena. The Coulomb potential, 
$v(r)=a_{-1}/r$, describes the interaction between charged particles. A Coulomb plus linear potential, $v(r)=a_{-1}/r+a_{1}r$, also known as Cornell potential, is the most common potential for modeling the confining quark interaction. It is also used in atomic physics to describe the Stark effect which occurs when the hydrogen atom is placed in an electric field. The two dimensional Coulomb plus quadratic potential, $v(r)=a_{-1}/r+ a_{2}r^{2}$, is related to the Zeeman effect; the hydrogen atom in magnetic field. The quartic harmonic oscillator potential, $v(r)=a_{2}r^{2}+a_{4}r^{4}$, is used in field theory to model the spontaneous  breaking of symmetry. It is evident that there is a great deal of physics that depends on the precise knowledge of the dynamics of the Coulomb potential with various polynomial interactions.

Over the years, several approaches have been developed to study some special cases of this problem (for a recent review see Ref.\ \cite{plante}). To the best of our knowledge, no method has been proposed yet that could treat this problem with arbitrary potential strength $a_{i}$ and arbitrarily high power of $k$.

In this work we calculate the Green's function of the non-relativistic quantum Hamiltonian with a Coulomb plus polynomial potential. The Green's function of polynomial potentials on the harmonic oscillator basis has been calculated before \cite{grechi,znojil}. The use of the Coulomb-Sturmian basis will allow us to incorporate the Coulomb potential.

If we know the Green's operator, then we have complete knowledge of the physical system. For example, the eigenvalues of the Hamiltonian coincide with the poles of the Green's operator.  The corresponding eigenvectors can be determined from the relation
\begin{equation}
|\psi_{n}\rangle \langle \psi_{n} | = \frac{1}{2\pi i} \oint_{C} G(z) dz,
\end{equation}
where $C$ encircles the eigenvalue $E_{n}$ in a counterclockwise direction without incorporating other poles.

In our previous works, Refs.\ \cite{klp,dhp}, the Coulomb Green's operator was calculated in the Coulomb-Sturmian basis. In this basis, the operator $J=z-H$ has an infinite symmetric tridiagonal, i.e. Jacobi, or J-matrix structure, where $z$ is a complex number. We have shown that the $G(z)=(z-H)^{-1}=J^{-1}$ Green's operator can be calculated in terms of continued fractions. This Coulomb Green's operator was used to solve the Faddeev integral equations of the three-body Coulomb problem \cite{fagyi}.

In the Coulomb-Sturmian basis, the Hamiltonian with  the potential  (\ref{v1}) is represented by an infinite symmetric band matrix. An infinite band matrix can always be considered as a block-Jacobi matrix with $m\times m$ blocks, where $m$ is finite. Thus the continued fraction becomes a matrix-valued continued fraction.

This paper is organized as follows. In Section II we introduce the $D$-dimensional Coulomb-Sturmian basis. In Section III we calculate the infinite band matrix representation of the Hamiltonian. In Section IV we derive the matrix continued fraction for the Green's operator. Some applications are considered in Section V. First, to demonstrate the power of the method we show an analytically known case, the harmonic oscillator in two and three dimensions. Then we consider the Coulomb plus linear confinement potential $v(r)=a_{-1}/r+a_{1}r$ in three dimensions and the Coulomb plus quadratic confinement potential, $v(r)=a_{-1}/r+ a_{2}r^{2}$, in two dimensions.

\section{The Coulomb-Sturmian basis}

The kinetic energy operator in $D$-dimension, with $ D\geq 2$, is given by 
\begin{equation}
H_{0} = - \frac{1}{2} \left( \frac{d^2}{d r^2} - \frac{  L(L+1)  }{r^{2}} \right),
\end{equation}
where $L = l +(D-3)/2$.
The Coulomb-Sturmian (CS) functions are the solutions of the Sturm-Liuoville problem of the Coulomb Hamiltonian  \cite{rotenberg}
\begin{equation}\label{SL}
\left( H_{0}    -   \frac{\lambda}{r}   \right)  \langle r | n \rangle =- \frac{b^{2}}{2}   \langle r | n \rangle,
\end{equation}
where $b>0$ is a parameter, $n$ is the radial quantum number, $n=0,1,\cdots$, and $\lambda=(n+L+1)b$.
In coordinate space, the CS functions are given by
\begin{equation}\label{eq:CS-basis}
\langle r |n \rangle =  \left[ \frac{\Gamma(n+1)}{\Gamma(n+2L+2)}\right]^{1/2}
\hbox{\rm e}^{-b r} (2 b r)^{L+1} L_n^{2L+1}(2 b r)~,
\end{equation}
where  ${L}^\alpha_{n}$ is an associated Laguerre polynomial. 
The CS functions form a basis. With $\langle r | \tilde{n} \rangle =1/r \langle r | n \rangle$, we have the orthogonality
\begin{equation}\label{orthog}
\langle \tilde{n} | n' \rangle =  \langle n | \tilde{n'} \rangle =    \langle n | 1/r | n' \rangle =\delta_{n n'},
\end{equation}
and the completeness relations
\begin{equation}\label{completeness}
\sum_{n=0} | n \rangle \langle \tilde{n} |= \sum_{n=0} | \tilde{n} \rangle \langle {n} | =
{\bf 1}.
\end{equation}

\section{Band-matrix representation}

We consider the Hamiltonian
\begin{equation}
H = H_{0} +\sum_{i=-1}^{k} a_{i}r^{i}
\end{equation}
on the CS basis. 

By utilizing the relations
\begin{equation}\label{lrec}
L_{n}^{\alpha}=L_{n}^{\alpha+1}-L_{n-1}^{\alpha+1}
\end{equation}
and
\begin{equation}\label{lagint}
\int_{0}^{\infty}  \exp(-x) x^{\alpha} L_{n}^{\alpha}(x) L_{n'}^{\alpha}(x) dx =
\frac{\Gamma(n+\alpha +1)}{\Gamma (n+1)} \delta_{n,n'}
\end{equation}
we can easily calculate the overlap of the CS states 
\begin{equation}\label{overlap}
\langle n | n' \rangle = \langle n' | n \rangle= \left\{
\begin{matrix}
\frac{ \displaystyle 1}{ \displaystyle b} (n+L+1)  & \textrm{for}~~n=n'~, \\
 - \frac{ \displaystyle 1}{ \displaystyle 2b} \sqrt{n'(n'+2L+1)} & \textrm{for}~~ n'=n+1~, \\
 0 & \textrm{for}~~ n'>n+1. 
\end{matrix}
\right.
\end{equation}
Then starting form Eq.\ (\ref{SL}), and using Eqs.\ (\ref{orthog}) and (\ref{overlap}), we can derive the CS matrix elements of the kinetic energy operator
 \begin{widetext}
\begin{equation}\label{h0me}
\langle n | H_{0} | n' \rangle =  \langle n' | H_{0} | n  \rangle =  \left\{
\begin{matrix}
\frac{ \displaystyle  b}{ \displaystyle  2  }(n+L+1)  & \textrm{for}~~n'=n~, \\
 \frac{  \displaystyle b}{ \displaystyle  4 } \sqrt{n'(n'+2L+1)} & \textrm{for}~~ n'=n+1~, \\
 0 & \textrm{for}~~n'>n+1~. 
\end{matrix}
\right.
\end{equation}
From Eqs.\ (\ref{orthog}), (\ref{overlap}) and (\ref{h0me}) it follows that the Hamiltonian $H=H_{0}+a_{-1}/r+a_{0}$  is tridiagonal on the CS basis.

By further utilizing Eqs.\ (\ref{lrec}) and (\ref{lagint}) we can also derive the following matrix elements 
\begin{equation}
\langle n | r | n' \rangle =  \langle n' | r | n \rangle = \left\{
\begin{matrix}
 \frac{\displaystyle  1 }{ \displaystyle 4b^{2}}(6{n}^{2}+2(L+1)(6n+2L+3))  & \textrm{for}~~n'=n~, \\
 -\frac{\displaystyle  1 }{ \displaystyle 2b^{2}} (2n'+2L+1)\sqrt{n'(n'+2L+1)} & \textrm{for}~~ n'=n+1~, \\
 \frac{\displaystyle  1}{ \displaystyle 4b^{2}} \sqrt{n'(n'-1)(n'+2L)(n'+2L+1)} & \textrm{for}~~ n'=n+2~, \\
 0 & \textrm{for}~~n'>n+2, \\
\end{matrix}
\right.
\end{equation}
and
\begin{equation}
\langle n | r^{2} | n' \rangle =  \langle n' | r^{2} | n  \rangle = \left\{
\begin{matrix}
 \frac{ \displaystyle 1 }{ \displaystyle 8b^{3}}
 \left[   (  ( (10n+2L+4)(n+2L+3)+ 
 9n(n-1))(n+2L+2)+n(n-1)(n-2))  \right]  & \textrm{for}~~n'=n~, \\
 -\frac{ \displaystyle 3 }{ \displaystyle 8 b^{3}} \left[  
 (4n'+2L)(n'+2L+2)+(n'-1)(n'-2)  \right]
 \sqrt{n'(n'+2L+1)} & \textrm{for}~~ n'=n+1~, \\
 \frac{ \displaystyle 3}{ \displaystyle 8b^{3}} (2n'+2L)\sqrt{n'(n'-1)(n'+2L+1)(n'+2L)} & \textrm{for}~~ n'=n+2~, \\
 -\frac{ \displaystyle 1}{ \displaystyle 8b^{3}} \sqrt{n'(n'-1)(n'-2)(n'+2L+1)(n'+2L)(n'+2L-1)} & \textrm{for}~~ n'=n+3~, \\
0 & \textrm{for}~~n'>n+3. \\
\end{matrix}
\right.
\end{equation}
\end{widetext}
Similarly, one can derive $\langle n | r^{k} | n' \rangle$ CS matrix elements for $k>2$ as well. If $k$ is finite, the Hamiltonian is an infinite symmetric band matrix, if $k=1$, it is a pentadiagonal, if $k=2$ it is a septadiagonal band matrix.

\section{Matrix continued fraction representation of the Green's operator}

The Green's operator $G$ is formally defined by the equation
\begin{equation}\label{eq:gf_def}
J(z) G(z) = G(z) J(z)= 1~,
\end{equation}
where $J(z)=z-H$ and $z$ is a complex number. On the CS basis, this takes the form
\begin{equation}\label{gdefm}
\sum_{i'} \langle i | J(z) | i' \rangle \langle \tilde{i'} |  G(z) | \tilde{i''} \rangle  = \delta_{i,i''}~.
\end{equation}

Now the operator $J(z)$ has an infinite symmetric band-matrix structure. An infinite symmetric band matrix can always be considered as a block-tridiagonal or block-Jacobi-matrix. So, Eq.\  (\ref{gdefm}) looks like
\begin{equation}\label{eq:tri-diag-HG}
\begin{split}
&
\left(\begin{matrix}
J_{0,0} & J_{0,1} & {0}    & { 0}       & \ldots \\
J_{1,0} & J_{1,1} & J_{1 2} & { 0}       & \ldots \\
{ 0}     & J_{2,1} & J_{2,2} & J_{2,3} & \ldots  \\
{ 0}     & { 0}       & J_{3,2} & J_{3,3} & \ldots  \\
\vdots & \ddots & \ddots & \ddots & \ddots  \\
\end{matrix}\right) \\
& \times 
\left(\begin{matrix}
G_{0,0} & G_{0,1} & G_{0,2} & G_{0,3} & \ldots \\
G_{1,0} & G_{1,1} & G_{1,2} & G_{1,3} & \ldots \\
G_{2,0} & G_{2,1} & G_{2,2} & G_{2,3} & \ldots  \\
G_{3,0} & G_{3,1} & G_{3,2} & G_{3,3} & \ldots  \\
\vdots & \dots & \dots & \ddots & \ddots  \\
\end{matrix}\right) \\
& =
\left(\begin{matrix}
{ 1} & { 0} & { 0}  & { 0} & \ldots \\
{ 0} & { 1} & { 0}  & { 0} & \ldots \\
{ 0} & { 0} & { 1} & { 0} & \ldots  \\
{ 0} & { 0} & { 0} & { 1} & \ldots  \\
\vdots & \ddots & \ddots & \ddots & \ddots  \\
\end{matrix}\right)~,
\end{split}
\end{equation}
where $J_{n,n'}$ and $G_{n,n'}$ are $m\times m$ block matrices, and the ${ 1}$'s and ${0}$'s are $m\times m$ unit and null matrices, respectively.

Just knowing the $N\times N$ upper left corner of the full Green's matrix is sufficient to determine physical quantities. Let us denote the corresponding $N\times N$ upper left corner block matrices by $J^{(N)}$, $G^{(N)}$ and 
$1^{(N)}$, respectively. 
If we multiply the $N\times \infty$ part of $J$ with the $\infty \times N$
part of $G$ we get the $N\times N$ block unit matrix $1^{(N)}$. 
The sum, due to the block tridiagonal form of $J$, is reduced to three block terms
\begin{equation}\label{eq:gf-eq-2solve}
J_{n,n-1}G_{n-1,n'} + J_{n,n}G_{n,n'} + J_{n,n+1}G_{n+1,n'} = \delta_{n,n'}~,
\end{equation}
where $n=0,1,..N$ and $n'=0,1,..N$. If $n<N$, 
only terms from $G^{(N)}$ are participating in the sum. For  $n=N$ an extra block matrix 
$G_{N+1,n'}$ outside the truncated subspace is needed:
\begin{equation}\label{eq:gf-eq-2solveN}
J_{N,N-1} G_{N-1,n'} + J_{N,N}G_{N,n'} + J_{N,N+1} G_{N+1,n'} = \delta_{N,n'}~.
\end{equation} 
We can formally eliminate this block by writing
\begin{equation}\label{eq:gf-eq-2solveN2}
\begin{split}
& J_{N,N-1} G_{N-1,n'} + \\ 
& \left[ J_{N,N} + J_{N,N+1} {G_{N+1,n'}}( G_{N,n'})^{-1} \right] G_{N,n'} = 
\delta_{N,m}~.
\end{split}
\end{equation}
This formal elimination of the elements outside of $G^{(N)}$ amounts to modifying the $J_{N,N}$ block of $J^{(N)}$.

We can calculate $G_{N+1,n'} (G_{N,n'})^{-1}$ from another relation:
\begin{equation}\label{eq:gf-eq-2solveNN}
J_{N+1,N} G_{N,n'} + J_{N+1,N+1}G_{N+1,n'} + J_{N+1,N+2} G_{N+2,n'} = 0~.
\end{equation}
By introducing the notation
\begin{equation}\label{eq:cf01}
C_{N+1} = -G_{N+1,n'}  (G_{N,n'})^{-1} (J_{N+1,N})^{-1} ~,
\end{equation}
Eq.\ (\ref{eq:gf-eq-2solveN}) can be rearranged as
\begin{equation}\label{cn-1}
C_{N+1} = \left( { J_{N+1,N+1} - J_{N+1,N+2}  C_{N+2}   J_{N+2,N+1} } \right)^{-1}~.
\end{equation}
 A repeated application of this relation 
results in a continued fraction with block matrices
\begin{widetext}
\begin{equation}\label{cncf}
C_{N+1}= \left( { J_{N+1,N+1} - J_{N+1,N+2}  
 \left( { J_{N+2,N+2} - J_{N+2,N+3}  (J_{N+3,N+3} -  \cdots )^{-1}   J_{N+3,N+2} } \right)^{-1} J_{N+2,N+1} } \right)^{-1}~.
 \end{equation}
\end{widetext}

This matrix continued fraction does not depend on the index $n'$ and the correction term to $J_{N,N}$ is the same for all $n'$. Therefore, we can write
Eq.\ (\ref{eq:gf-eq-2solveN2}) in the form
\begin{equation}
( {J}^{(N)}_{i,j} - \delta_{i,N} \delta_{j,N} J_{N,N+1} C_{N+1} J_{N+1,N} ) G^{(N)}=1^{(N)},
\end{equation}
i.e.\ the modified $N\times N$ block-Jacobi matrix is the inverse of $G^{(N)}$
\begin{equation}\label{g000}
(G^{(N)})^{-1}=  {J}^{(N)}_{i,j} - \delta_{i,N} \delta_{j,N} J_{N,N+1}  C_{N+1} J_{N+1,N} ~.
\end{equation}

The numerical evaluation of matrix continued fractions is very similar to those of ordinary continued fractions. In backward evaluation we start at some $K>N$ term, neglect the higher terms, and evaluate (\ref{cncf}) from the 
inside out. If a new approximant with larger
$K$ is needed, we have to start the whole process again. On the other hand, the backward evaluation is simple and provides very accurate results.

\section{Examples}

To demonstrate the power of this method, we take first the harmonic oscillator
\begin{equation}\label{hoh}
H=H_{0}+{1}/{2}\; \omega^{2}r^{2}
\end{equation}
in two and three dimensions. This Hamiltonian has a septadiagonal structure on the CS basis, which can be considered as a block-Jacobi matrix with $3\times 3$ blocks. In our numerical example we take $\omega=1$ and roll up the continued fraction up to the first block and calculate the determinant of a $3 \times 3$ matrix. Figure 1 shows the 
poles of $G^{0}(z)$ as function of the CS parameter $b$. We took
$l=0$, which implies $L=0$ for $D=3$ and $L=-1/2$ for $D=2$, respectively. 
It can be seen that even a $3\times 3$ Green's matrix provides all the eigenvalues of (\ref{hoh}). They agree with the exact results $E_{n}=\omega(2n+L+3/2)$ up
to machine accuracy and  the results are independent of the choice for the parameter $b$ of the CS basis. 

\begin{figure}[tbp]
\begin{center}
\includegraphics[width=8.cm]{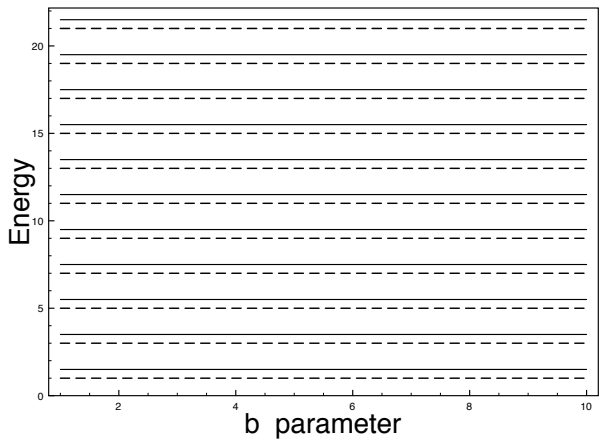}
\caption{Eigenvalues of the three-dimensional (full line) and the two-dimensional (broken line) harmonic oscillator as a function of the CS basis parameter $b$.}
\label{table1}
\end{center}
\end{figure}

Next, we consider the Coulomb plus linear potential in three dimensions
\begin{equation}\label{cornell}
H=H_{0} + Z/r + \alpha r,
\end{equation}
and the Coulomb plus quadratic potential in two dimensions
\begin{equation}\label{cr2}
H=H_{0} + Z/r + {1}/{2}\; \omega^{2} r^{2}.
\end{equation}
The Hamiltonian (\ref{cornell}) is pentadiagonal, and can be considered as a block-Jacobi matrix with $2\times 2$ blocks, while (\ref{cr2}) is septadiagonal, like in the harmonic oscillator case.
For the numerical values we take $Z=-1$, $\alpha=1$, and $\omega=1$. 
The lowest $20$ eigenvalues are given in Table I.
We  observed a similar stability with respect to varying $b$ as we did in the case of the harmonic oscillator.

\begin{table}[h]
\begin{center}
\caption{Eigenvalues of Hamiltonians with Coulomb plus linear potential in three dimensions and Coulomb plus quadratic potential in two dimensions.  The potential parameters are
$Z=-1$, $\alpha=1$ and $\omega=1$.}
\begin{tabular}{|c|c|c|} \hline  
 $n$ & Eq.\ (\ref{cornell}) & Eq.\ (\ref{cr2}) \\ \hline 
 1 & 0.577921351961 &  -1.836207439051 \\
 2 & 2.450162895052 & 1.576895542024 \\ 
 3 & 3.756905691262 & 3.828388290161 \\
 4 & 4.855671243373 &  5.963137645126  \\
 5 & 5.836029886654 & 8.052626115348 \\
 6 & 6.736620996511 & 10.11839697526 \\
 7 & 7.578378030294 & 12.16972896261 \\
 8 & 8.374205689360 & 14.21142722055 \\
 9 & 9.132754730978 &   16.24628453060  \\
 10 & 9.860176266906 & 18.27605894134 \\
 11 & 10.56103960914 &  20.30192413905  \\
 12 & 11.23885563715 &   22.32469992791 \\
 13 & 11.89639544211 & 24.34497987508 \\
 14 & 12.53589461658 & 26.36320650647 \\
 15 & 13.15918982353 & 28.37971786276 \\
 16 & 13.76781330561 &  30.39477752867 \\
 17 & 14.36306021727 &  32.40859467947 \\
 18 & 14.94603779901 &  34.42133786062 \\
 19 & 15.51770206715 & 36.43314470188\\
 20 & 16.07888570444 &  38.44412891767 \\
 \hline
\end{tabular}
\end{center}
\label{table}
\end{table}

\section{Summary and conclusions}

In this work, we have shown that in the $D$-dimensional Coulomb-Sturmian basis the non-relativistic $D$-dimensional Hamiltonian with Coulomb plus polynomial potential has a band-matrix structure. A band matrix can always be considered as a block-Jacobi matrix, and, consequently, the Green's matrix can be constructed in terms of matrix continued fractions. A numerically converged matrix continued fraction gives a numerically exact Green's operator, which even on a very small basis provides the exact spectrum. We have demonstrated the power of the method in the case of harmonic oscillator and obtained the exact spectrum.

As examples, we studied the Coulomb plus linear confinement in three dimensions and the Coulomb plus quadratic confinement in two dimensions. The first is related to the quark confinement and the Stark effect, while the second one is related to the Zemann effect. The exact knowledge of these Green's operators may facilitate the use of integral equations to describe quantum processes in external fields.

\begin{acknowledgments}
This work has been supported by the Research Corporation.
\end{acknowledgments}


\begin{thebibliography}{99}

\bibitem{plante} G.~Plante and A.~F.~Antippa,  J.\ Math.\ Phys.\  {\bf 46}, 062108 (2005).

\bibitem{grechi} S.\ Graffi and V.\ Grecchi, Lett.\ Nuovo Cimento {\bf 12}, 425 (1975).

\bibitem{znojil} M.\ Znojil and L.\  Majling, J.\ Phys A: Math. Gen. {\bf 16} 639 (1983).

\bibitem{dhp} F.\ Demir, Z.\ T.\ Hlousek, and Z.\ Papp,  Phys. Rev. A {\bf 74}, 014701 (2006).

\bibitem{klp} B.\ K\'onya, G.\ L\'evai, and Z.\ Papp,  J.\ Math.\ Phys. {\bf 38}, 4832 (1997).

\bibitem{fagyi} See eg.   Z.\ Papp, Phys.\ Rev.\ C, {\bf 55}, 1080 (1997);
 Z.\ Papp,  C-.Y.\ Hu, Z.\ T.\ Hlousek, B. K\'onya and S.\ L.\ Yakovlev,
Phys.\ Rev.\ A,  {\bf 63}, 062721 (2001); Z.\ Papp, J.~Darai, J.~Zs.~Mezei, Z.~T.~Hlousek, and  C-.Y.\ Hu, 
Phys. Rev.\ Lett.\ {\bf 94}, 143201 (2005).

\bibitem{rotenberg} M.\ Rotenberg, Adv.\ At.\ Mol.\ Phys.\ {\bf 6}, 233 (1970).


\end{thebibliography}
\end{document}